\documentclass[%
 aip,
 jcp,%
 amsmath,amssymb,
 preprint%,
]{revtex4-1}

\usepackage{graphicx,color}
\graphicspath{{figure/}}
\usepackage{subfig}
\usepackage{dcolumn}% Align table columns on decimal point
\usepackage{bm}% bold math
\providecommand{\noopsort}[1]{}\providecommand{\singleletter}[1]{#1}%
\begin{document}

%\preprint{AIP/123-QED}

\title[]{Pathways of mechanical unfolding of $FnIII_{10}$: low force intermediates}%\footnote{Error!}}% Force line breaks with \\
%\thanks{Footnote to title of article.}

\author{M. Caraglio}
\email{michele.caraglio@polito.it}
\affiliation{Dipartimento di Fisica and CNISM, Politecnico di Torino, c. Duca degli Abruzzi 24, Torino, Italy}
\affiliation{INFN, Sezione di Torino, Torino, Italy}
\author{A. Imparato}
\email{imparato@phys.au.dk}
\affiliation{Department of Physics and Astronomy, University of Aarhus,
  Ny Munkegade, Building 1520, DK--8000 Aarhus C, Denmark}
\author{A. Pelizzola}
\email{alessandro.pelizzola@polito.it}
\affiliation{Dipartimento di Fisica and CNISM, Politecnico di Torino, c Duca degli Abruzzi 24, Torino, Italy}
\affiliation{INFN, Sezione Torino, Torino, Italy}

\date{\today}% It is always \today, today,

\begin{abstract}
  We study the mechanical unfolding pathways of the $FnIII_{10}$
  domain of fibronectin by means of an Ising--like model, using both
  constant force and constant velocity protocols. At high forces and high velocities our
  results are consistent with experiments and previous
  computational studies. Moreover, the simplicity of the model allows us to probe the
  biologically relevant low force regime, where we predict the
  existence of two intermediates with very close elongations. The
  unfolding pathway is characterized by stochastic transitions between
  these two intermediates.
\end{abstract}

\pacs{Valid PACS appear here}% PACS, the Physics and Astronomy
\keywords{Suggested keywords}%Use showkeys class option if keyword
\maketitle

\section{Introduction}

The mechanical unfolding of biopolymers has been the subject of an
intense research activity, both experimental and theoretical, in the
last two decades. For a recent review, see \cite{KumarLi}. Innovative 
single molecule experimental techniques, mainly based on atomic force
microscopy (AFM) and optical tweezers, have been used to investigate the response of biopolymers to controlled forces,  while theoretical and computational models
at different levels of coarse graining have been proposed and
investigated.

Among the various molecules studied, fibronectin is particularly
important, due to its role in tissue elasticity, cell adhesion and
cell migration \cite{Geiger2001}. Its 10th type III module ($FnIII_{10}$) is known to be
crucial for cell adhesion, through the binding of its RGD motif to
transmembrane integrin receptors. The secondary structure of this
module consists of 2 antiparallel $\beta$--sheets forming a
$\beta$--sandwich. The $\beta$--strands are usually denoted with
letters from A (the strand closest to the N terminal) to G (the C
terminal one). The two sheets are made of strands ABE and DCFG,
respectively, and the RGD motif is in the loop separating strands F
and G.

\begin{figure}[htbp]
\centering
\includegraphics[width=0.45\textwidth]{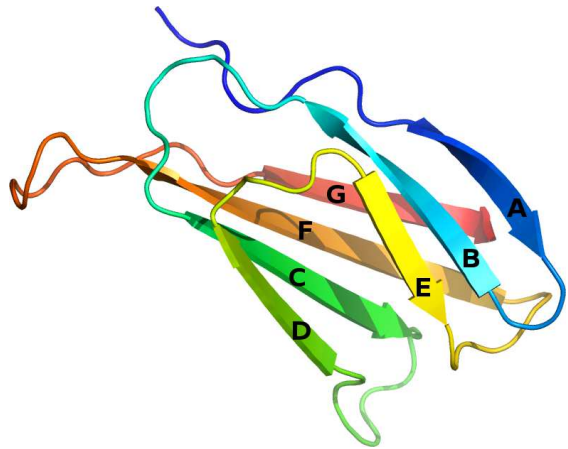}
\caption{Sketch of the native structure of $FnIII_{10}$ (Protein Data Bank ID 1ttf) with $\beta$--strands labeled A--G in sequence order. Figure generated by PyMOL. }
\label{fig:fibronectin3D}
\end{figure}

The mechanical unfolding of $FnIII_{10}$ has been studied both
experimentally \cite{JMolB319,JMolB345} and by computer simulations
\cite{ProcNatlAcadSciUSA96,PaciKarplus,ProcNatlAcadSciUSA97,JMB323,BiophysJ96}. Single
molecule AFM experiments have shown that $FnIII_{10}$ has a low
mechanical stability, compared to other fibronectin type III domains
\cite{JMolB319}. Furthermore, AFM experiments by the same group
\cite{JMolB345} showed  that $FnIII_{10}$ can unfold according
to different pathways. Apparent two--state transitions were observed,
as well as unfolding through intermediate states. Experiments on
suitable mutants suggested the possible existence of two different
intermediate states, which is also consistent with some simulations
\cite{PaciKarplus,JMB323}, while other simulations predicted simpler
\cite{ProcNatlAcadSciUSA96,ProcNatlAcadSciUSA97} or more complex
\cite{BiophysJ96} scenarios. 

In the present paper we shall study the mechanical unfolding of
$FnIII_{10}$ by means of a generalized Ising--like model we have
recently proposed \cite{PRL98,JCHEMPHYS127,PRL100,PRL103}. The model
has already been shown \cite{PRL98,JCHEMPHYS127} to reproduce the
general features of mechanical unfolding experiments, like the force
dependence of the average unfolding time in a constant force protocol,
or the rate dependence of the unfolding force in a constant rate
protocol, together with the corresponding probability
distributions. The same model turned out to predict the correct values
 for the
unfolding lengths of a titin domain \cite{PRL98,JCHEMPHYS127} and of
ubiquitin \cite{PRL100}. Moreover, it has been used to investigate the
unfolding pathways of ubiquitin \cite{PRL100} and of a 236--base RNA
fragment \cite{PRL103}, and the resulting pathways  turned out to be 
consistent with both experimental  and computational results, where 
more detailed molecular models were used.

In the case of $FnIII_{10}$ we are particularly interested in
exploring the biologically relevant low force regime
\cite{Erickson,Geiger2001,Li2002}, which is thought to be close to the
equilibrium unfolding force and cannot be explored by simulations of
more detailed, and more computationally expensive, molecular
models. Our model can probe forces close to the equilibrium unfolding
force, whose value we use to set our force unit. Such value is
unfortunately not exactly known. Erickson \cite{Erickson} estimates
the equilibrium unfolding force to be at most 5 pN, on the basis of an
order of magnitude calculation. On the other hand, an estimate close
to 20 pN was reported in \cite{BiophysJ96}. Choosing the value of 20
pN, in order to set our force unit, our results give unfolding forces
in very good agreement with the AFM experiments (see Sec.\
\ref{Pathways}).

The paper is organized as follows: in Sec.\ \ref{ModelMethods} we
shall describe our model and simulation techniques; some equilibrium
results are discussed in Sec.\ \ref{Equilibrium}; in Sec.\
\ref{Pathways} we shall describe our results for the unfolding
pathways, using both constant force and constant velocity protocols;
finally, in Sec.\ \ref{Conclusions} we shall draw some conclusions. 

\section{Model and Methods}
\label{ModelMethods}

\subsection{Model}

We represent the polypeptide chain pulled by an external force through a simple Ising-like model \cite{PRL98,JCHEMPHYS127,PRL100,PRL103}, that is a generalization of the Wako-Sait\^{o}-Mu\~{n}oz-Eaton (WSME) G\={o}--type model for protein folding \cite{JPhysSocJpn441931,JPhysSocJpn441939,Nature390,ProcNatlAcad95,ProcNatlAcad96,Science298,AmosEPL,PRL88,JSTAT2005,PRL97,JSTAT2006,PRL99,JCHEMPHYS126}. 

In this model a $\left( N+1 \right) $ residues polypeptide chain is described by $N$ binary variables $m_k$, $\left( k=1,\ldots,N \right) $, associated to the peptide bonds. The variable $m_k$ is equal to $1$ if the $k$-th bond is in a native state and $0$ if the bond is in an unfolded state. % The $N$-tuple $m = \left\lbrace m_k \right\rbrace  \in \left\lbrace 0,1 \right\rbrace ^N$ completely describes the state of the molecule.

In the following a {\it native stretch} will indicate  a sequence of consecutive amino acids connected by native bonds and delimited by two non-native bonds. 
In order to characterize the state of such a stretch, we introduce the quantity
\begin{equation}
S_{ij} \equiv \left( 1-m_i \right) \left( \prod_{k=i+1}^{j-1} m_k \right) \left( 1-m_j \right) \; ,  
\end{equation}
with $\left( 0 \leqslant i < j \leqslant N+1 \right) $ and the boundary conditions $m_0=m_{N+1}=0$. 
Thus, a native stretch delimited by bonds $i$ and $j$ is characterised by $S_{ij}=1$, while 
if the sequence  is not a native stretch, $S_{ij}=0$. In the case of $j=i+1$ the stretch corresponds to the single $(i+1)$-th residue. 
Therefore, given a certain configuration $m = \left\lbrace m_k \right\rbrace$ of the chain, the number $M$ of native stretches is equal to
$ M = 1+ \sum_{i=1}^N \left( 1-m_i \right) $.

%In this picture the number $M$ of stretches is equal to $1$ plus the number of $0$'s in the $N$-tuple $m = \left\lbrace m_k \right\rbrace  \in \left\lbrace 0,1 \right\rbrace ^N$, that is $ M = 1+ \sum_{i=1}^N \left( 1-m_i \right) $.

In the WSME model two amino acids can interact only if they are in
contact (i.e. if they have at least a pair of nonhydrogen atoms closer
than $4$ \AA\ in the native structure deposited in the Protein Data
Bank (PDB)) and if they belong to the same stretch. The effective
Hamiltonian reads:
\begin{equation} \label{HWSME}
H \left( m,q \right) = \sum_{i=1}^{N-1} \sum_{j=i+1}^{N} \epsilon_{ij} \Delta_{ij} \prod_{k=i}^{j} m_k - k_B T \sum_{i=1}^{N} q_i(1-m_i) \; ,
\end{equation}
where $\epsilon_{ij}$ is the energy gain associated to the contact between $i$-th and $(j+1)$-th amino acids (see next subsection for details), $\Delta$ is called contact matrix and its $(i,j)$ element takes the value $\Delta_{ij} = 1$ if such a contact exists in the native structure while $\Delta_{ij} = 0$ otherwise or if $j = i+1$ (since if the amino acids are so close in the chain sequence, they have pair of atoms closer than $4$ \AA\ even in the unfolded state). The quantity $q_i > 0$ is the entropic cost of ordering bond $i$. In the following we define $h_{ij} = \epsilon_{ij} \Delta_{ij}$. In Ref.\cite{PRL88} it has been discussed how to compute exactly the partition function of the WSME model through a transfer-matrix formalism.

\medskip

In our generalized model we substitute the entropic term for the coupling to the force described by a potential energy function $V$ depending on the end-to-end length of the protein $\mathcal{L}$:
\begin{equation}
H \left( m, \mathcal{L} \right) = \sum_{i=1}^{N-1} \sum_{j=i+1}^{N} h_{ij} \prod_{k=i}^{j} m_k + V \left( \mathcal{L} \right)   \; .
\end{equation}
In our simulations the protein is pulled either by a constant force or at constant pulling velocity. In the first case the force $f$ applied to the protein ends is constant and the potential energy takes the form $V \left( \mathcal{L} \right) = -f \mathcal{L}$, while in the second case the energy is time--dependent: $V \left( \mathcal{L} \right) = \frac{k}{2} \left( \mathcal{L}_0 + vt - \mathcal{L} \right) ^2 $, where $k$ is a spring constant, $v$ is the pulling velocity and $\mathcal{L}_0$ is the initial equilibrium elongation.

In order to define the length $\mathcal{L}$ we assume that each stretch ($S_{ij}=1$) can be only parallel or antiparallel to the direction of the applied force and we implement such an assumption through a new binary variable $\sigma_{ij}$ that can take the values $+1$ or $-1$ respectively. The end-to-end length of the protein is defined as the sum of the lengths $l_{ij}$ of each stretch multiplied by $\sigma_{ij}$:
\begin{equation}
\mathcal{L}(m,\sigma) = \sum_{i=0}^N \sum_{j=i+1}^{N+1} l_{ij} \sigma_{ij} S_{ij} \; ,
\end{equation}
where the dynamic variable $\sigma$ is a set of $M$ variables $\sigma_{ij}$, each one associated to a stretch. Since the $i$-th aminoacid is represented by the sequence of its $N_i$ nitrogen, $C_{\alpha , i}$ central carbon and $C_i$ carbon atoms, the lengths $l_{ij}$ are obtained from the PDB structure as the native distance between the midpoint of the $C_{i-1}$ and $N_i$ atoms and the midpoint of the $C_j$ and $N_{j+1}$ atoms.

Dealing with the case of constant force, since the variables $\sigma_{ij}$ do not interact among themselves, it is possible to obtain an effective Hamiltonian which has the same structure of the Hamiltonian (\ref{HWSME}) of the initial model and therefore the equilibrium thermodynamics is exactly solvable also in this case. In fact, given the Hamiltonian:
\begin{equation}
H \left( m, \sigma ; f \right) = \sum_{i=1}^{N-1} \sum_{j=i+1}^{N} h_{ij} \prod_{k=i}^{j} m_k -f  \sum_{i=0}^N \sum_{j=i+1}^{N+1} l_{ij} \sigma_{ij} S_{ij}  \; ,
\end{equation}
we can perform the sum on the $\sigma$ variables in the partition function:
\begin{equation} \label{eq:Z(f)}
\mathcal{Z} (f) = \sum_{ \left\lbrace m \right\rbrace } \sum_{ \left\lbrace \sigma \right\rbrace } e^{-\beta H \left( m,\sigma ;f  \right)} = \sum_{ \left\lbrace m \right\rbrace } e^{-\beta H_{\mathtt{eff}} \left( m; f  \right)} \; ,
\end{equation}
with 
\begin{equation}
H_{\mathtt{eff}} \left( m; f  \right) =  \sum_{i=1}^{N-1} \sum_{j=i+1}^{N} h_{ij} \prod_{k=i}^{j} m_k - \frac{1}{\beta} \sum_{i=0}^N \sum_{j=i+1}^{N+1} \ln \left[ 2 \cosh \left( \beta f l_{ij} \right) \right] S_{ij} \; .
\end{equation}
In the case of force $f=0$ the last expression reduces to eq.~(\ref{HWSME}) with $q_i = \ln 2$ for every $i$.

\subsection{Model parameters, simulation and analysis}

The parameters $\epsilon_{ij}$, as in \cite{ProcNatlAcad96,PRL88,JCHEMPHYS126,PRL98,JCHEMPHYS127,PRL100}, are taken equal to $n \epsilon$ where $n$ is an integer such that $5(n-1)<n_{at} \leqslant 5n$, and $n_{at}$ is the number of pairs of atoms in contact (that is, closer than $4$ \AA) between $i$-th and $(j+1)$-th amino acids. The temperature is set to $T = 0.768\, T_m$, where $T_m$ is the equilibrium unfolding temperature at zero force. Since experimentally $T_m = 375$ K \cite{Ingham}, we have $T = 288$ K. The force unit is then set in such a way that the equilibrium typical unfolding force at $T=288$ K is 20 pN. Since  an experimental measurement of this quantity is missing, it has been chosen on the basis of the estimates reported in \cite{BiophysJ96}. A detailed discussion about the choice of energy and force scales in the model has been reported in \cite{PRL100}.

The nonequilibrium unfolding kinetics have been studied by Monte Carlo (MC) simulations. More precisely, in the framework of a master equation approach \cite{PRL97}, we choose transition rates according to the Metropolis algorithm. Rigorously speaking, this choice cannot be derived from an underlying microscopic dynamics of the molecule. Nevertheless, it has been shown \cite{PRL98,JCHEMPHYS127,PRL100,PRL103} that it reproduces many quantitative and qualitative aspects of folding and unfolding of real molecules under an external force. A single MC step consists of a single--bond flip on the variable $m_j$, chosen with equal probability among the $N$ peptide bond variables, followed by a single--spin flip on the variable $\sigma_{ij}$, also chosen with uniform probability among the $M$ stretch orientational variables \cite{tesiMarcoZ}. In Sec.\ \ref{Pathways}, by comparing our estimated zero--force unfolding time with the corresponding experimental value, we shall find that a MC step corresponds to about 25 ns. 

Simulations have been run with nine values of the force ($122$ pN, $98$ pN, $81$ pN, $65$ pN, $53$ pN, $46$ pN, $40$ pN, $36$ pN, $28$ pN) and six constant pulling velocities (0.03 $\mu$m/s, 0.05 $\mu$m/s, 0.1 $\mu$m/s,0.3 $\mu$m/s, 0.5 $\mu$m/s, 1 $\mu$m/s), for each value of the force or of the velocity $100$ different unfolding trajectories have been considered. 

Each simulation stops $10^5$ MC steps after the protein reaches the value $ L_{u} = \frac{1}{2} L_{max} = \frac{1}{2} \sum_{i=0}^{N} l_{i,i+1} $. An exception is the case $f=28$ pN where we take $ L_{u} = \frac{2}{3} L_{max}$, because of the larger length fluctuations and in order to prevent the trajectory ending before a complete unfolding event takes place. 

In order to trace unfolding pathways,  we use %, as in \cite{PRL103}, 
the weighted fraction of native contacts as order parameter:
\begin{equation}
\phi_s = \dfrac{\sum_{i=r_1(s)}^{r_2(s)-2} \sum_{j=i+1}^{r_2(s)-1} h_{ij} \prod_{k=i}^{j} m_k}{\sum_{i=r_1(s)}^{r_2(s)-1} \sum_{j=i+1}^{r_2(s)} h_{ij}} \; ,
\label{eq:fnc}
\end{equation}
where $s$ is the string of bonds we are analysing and $r_1(s)$, $r_2(s)$ its first and last peptide units. As an example the string containing strands A and B has $r_1(AB)=6$ and $r_2(AB)=23$. A straightforward generalization is necessary for order parameters of strings of non-consecutive strands (i.e. C-F and B-E). $\phi_s$ turns out to be a better order parameter than the fraction of native bonds used in previous works because of its greater stability with respect to fluctuations. When discussing the folded or unfolded character of an individual $\beta$--strand, appropriate order parameters can be identified on the basis of the secondary structure. As an example, strand F appears in a $\beta$--sheet between strands C and G, which suggests to use $\phi_{CF}$ and $\phi_{FG}$ as order parameters for strand F. 

%Whether an individual $\beta$-strand is still folded \alb{($\phi_s\simeq1$)} or not \alb{($\phi_s\simeq0$)} is determined on the basis of the order parameter measured on the $\beta$-strands pairs it forms with other strands \alb{non capisco cosa vuoi dire??}.

\section{Equilibrium properties}
\label{Equilibrium}

As mentioned before, the equilibrium thermodynamics can be solved exactly in our model and we can thus follow the macroscopic state behaviour of the protein at different pulling forces. In \cite{epaps}  the average fraction of native bonds  and end--to--end length are plotted as functions of the force $f$.

\begin{figure}[htbp]
\centering
\includegraphics[angle=270, width=0.45\textwidth]{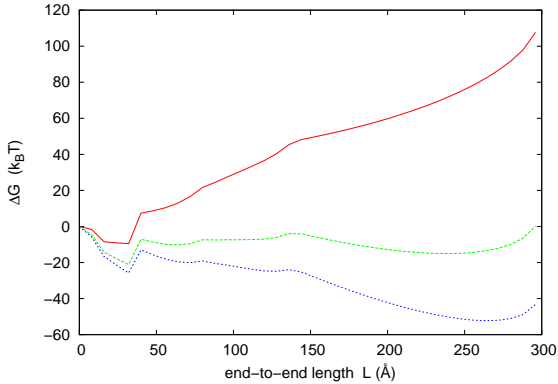}
\caption{Free energy landscape at temperature $T=288$ K and for forces $f = 0$ pN (red line), $f = 20$ pN (green line) and $f = 28$ pN (blue line). $\Delta G= G(L)-G(0)$.}
\label{fig:Elandscape}
\end{figure}

To obtain the equilibrium energy landscape of the protein as a function of the reaction coordinate $L$ we expand, following \cite{PRL88,JCHEMPHYS127}, the partition function~(\ref{eq:Z(f)}) in powers of $e^{\beta f}$ as
$$ \mathcal{Z} (f) = \sum_{ L=-L_{\mathtt{max}}}^{ L=L_{\mathtt{max}}} Z_0(L) \, e^{\beta f L} \; , $$
where $Z_0(L)$ is a zero force partition function constrained at the length value $L$:
$$ Z_0(L) = \sum_{m} \sum_{\sigma} \delta(L-L(m,\sigma))\, e^{-\beta H(m,\sigma;f=0)} $$
which, as $\mathcal{Z} (f)$, can be computed exactly.
The corresponding free energy reads:
$$ F_0(L) =  -k_BT \ln Z_0(L) \, .$$
In presence of a constant force, the free energy landscape is tilted and is given by $G(L) = F_0(L)-fL$. Fig.\ \ref{fig:Elandscape} shows the landscape for various forces: at zero force there is just one minimum at about 3.5 nm corresponding to the folded state. By increasing the force three more minima appear: two of them (end--to--end lengths of about 6 and 13 nm) are always local minima, and will be later associated to intermediate states, while the third one, corresponding to the fully unfolded state, becomes the global minimum when the force exceeds   20 pN.

\section{Unfolding pathways}
\label{Pathways}

\subsection{Force Clamp}

In the force clamp protocol the molecule is first equilibrated in absence of force, then at $t = 0$ the force instantaneously jumps to a non--vanishing constant value, which ranges between 28 to 122 pN. Notice that the forces we use are much closer to the equilibrium unfolding force, and hence to in vivo conditions, than most previous works, since more detailed models can be simulated only for very short time intervals. The smallest force probed by Karplus and Paci \cite{PaciKarplus} was 69 pN, and they did not observe any unfolding event at this force, while Gao et al \cite{JMB323} used forces not smaller than 400 pN. Only in the all--atom Monte Carlo simulations by Mitternacht et al \cite{BiophysJ96}  unfolding events at constant forces as small as 50 pN could be observed.

\begin{figure}[htbp]
\centering
\includegraphics[angle=270, width=0.45\textwidth]{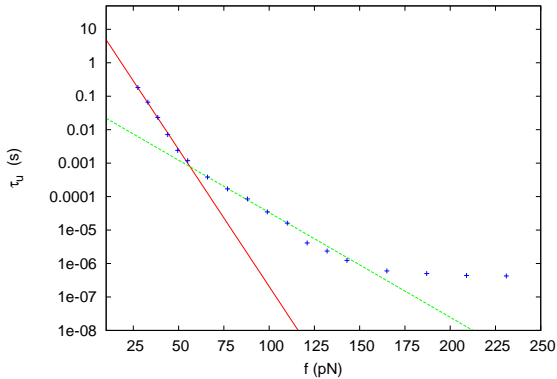}
\caption{Mean unfolding time $\tau_u$ as a function of the force $f$ applied to the molecule (average over 100 different trajectories). The red line is a fit to the Arrhenius' law in the range of forces from $25$ to $60$ pN. In this range we find from the fit $x_u = 3.4 \pm 0.1$ \AA . The green line is a fit from $60$ to $115$ pN, $x_u = 1.3 \pm 0.1$ \AA . }
\label{fig:unftime_vs_force}
\end{figure}

In Fig.\ \ref{fig:unftime_vs_force} we report the average unfolding time $\tau_u$ as a function of force $f$. Three regimes are clearly distinguishable. In the high force regime the unfolding time  saturates to a constant plateau, as observed for several other proteins \cite{KumarLi}. In the low force regime (25 to 60 pN) we have made a fit to the Arrhenius' law 
$$ \tau_u = \tau_0 \exp \left[ - \dfrac{fx_u}{k_B T} \right] \; ,$$ 
obtaining the unfolding length $x_u = 3.4 \pm 0.1$ \AA, which compares well, given the extreme simplicity of our model, with the experimental results $x_u = 3.8$ \AA\ \cite{JMolB319}. Comparing our zero--force unfolding time $\tau_0$ with its experimental value $\tau_{\mathtt{exp}} = 50 \, \mbox{s}$ \cite{JMolB319}, we find out that a single MC step in our model corresponds to about $25$ ns. In the intermediate force regime (60 to 115 pN) our fit yields $x_u = 1.3 \pm 0.1$ \AA.

The unfolding trajectories can be grouped in four classes according to their main features, i.e. their end-to-end length plateaus (if they exist) and the order parameters behaviour for the whole molecule and its various pairs of $\beta$-strands. 

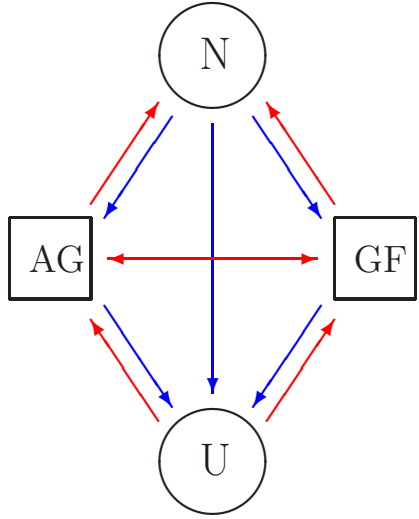
\begin{figure}
\setlength{\unitlength}{0.9cm}
\begin{center}
\begin{picture}(7,8)(0,0)
\thicklines
\put(0,0){\makebox(7,8){}}

\put(3.5,1){\circle{3}}
\put(3.5,7){\circle{3}}
\put(0.5,3.4){\line(1,0){1.2}}
\put(0.5,3.4){\line(0,1){1.2}}
\put(1.7,4.6){\line(0,-1){1.2}}
\put(1.7,4.6){\line(-1,0){1.2}}
\put(5.3,3.4){\line(1,0){1.2}}
\put(5.3,3.4){\line(0,1){1.2}}
\put(6.5,4.6){\line(0,-1){1.2}}
\put(6.5,4.6){\line(-1,0){1.2}}

\put(3.1,0.8){\begin{Large} U \end{Large}}
\put(3.1,6.8){\begin{Large} N \end{Large}}
\put(0.6,3.8){\begin{large} AG \end{large}}
\put(5.4,3.8){\begin{large} GF \end{large}}

\color{blue}

\put(3.5,6){\vector(0,-1){4}}
\put(2.9,6.1){\vector(-2,-3){1}}
\put(4.1,6.1){\vector(2,-3){1}}
\put(1.9,3.3){\vector(2,-3){1}}
\put(5.1,3.3){\vector(-2,-3){1}}

\color{red}
\put(1.95,4){\vector(1,0){3.1}}
\put(5.05,4){\vector(-1,0){3.1}}
\put(1.7,4.8){\vector(2,3){1}}
\put(5.3,4.8){\vector(-2,3){1}}
\put(4.3,1.6){\vector(2,3){1}}
\put(2.7,1.6){\vector(-2,3){1}}

\end{picture}
\end{center}
\setlength{\unitlength}{1cm}
\caption{\label{ClampScheme} Unfolding pathways scheme of $FnIII_{10}$ pulled by a constant force. Transitions denoted by red arrows have been observed only at low forces ($40, 36$ and $28$ pN). Oblique red arrows represent refolding transitions.}
\end{figure}

At large forces we observe simple 2--state trajectories, while at smaller forces various intermediates are obtained. A scheme of the possible pathways is shown in Fig.\ \ref{ClampScheme}. 

%\textit{In order to decide whether a plateau in the end-to-end length signals an intermediate or not we require that it must last at least for $93$ MC steps, i.e. the number of peptide units of the molecule. Thus, reasonably, the plateau is not just the time spended by the MC simulation to find a bond flip favourable for unfolding.}

In trajectories exhibiting intermediate states it turns out, as already pointed out in previous papers~\cite{ProcNatlAcadSciUSA96,ProcNatlAcadSciUSA97}, that strand G is always the first to break away. In cellular environment such behaviour seems to be connected to the function of the RGD motif Arg$78$-Gly$79$-Asp$80$~\cite{ProcNatlAcadSciUSA96,JMB323}. When the module is fully folded, the RGD motif is available for adhesion, while if strand G is pulled and detached from the remainder of the module, the RGD motif gets closer to the surface of the module and is not functional. 

%\textit{but in our simplified model these three residues don't play any particular role compare to the others in the $8$ residues long turn between strands F and G and and such event is due to the fact that the coupling of strand G to strand F is weaker than coupling between A and B. In fact it is reasonable that the molecule should start to unfold breaking away its terminal strands (i.e. A or G), but a naive consideration based only on summing the parameters $h_{ij}$ between various pairs of strands in the native structure $$ n(s) = \sum_{i=r_1(s)}^{r_2(s)-1} \sum_{j=i+1}^{r_2(s)} h_{ij} \; ,$$ with the results $n(AB) \sim 30$ and $n(FG) \sim 20$, gives to G a probability to detach that is more than $10^3$ times higher than that associated to A.}

%\textit{mancano considerazioni su termini entropici}

Strand G detachment may be rapidly followed by complete unfolding or by an intermediate state. A possibility is that strand A detaches almost at the same time of strand G while the remaining part of the molecule stays folded for a certain time before complete unfolding. This kind of unfolding pathway will be labeled with \textit{AG}, its intermediate end-to-end length is about $13.5$ nm. It may happen that instead of strand A, strand F detaches together with G, such unfolding pathway (intermediate end-to-end length $\sim14$ nm) will be labeled \textit{GF}.

The last possibility occurs only in the biologically relevant regime of low forces. It is believed \cite{BiophysJ96} that such relevant forces, in vivo, are of the same order of magnitude as the equilibrium unfolding force ($\sim 20$ pN, see section \ref{Equilibrium}), though forces as low as 5 pN have been suggested \cite{Erickson} as typical unfolding forces. Our low force unfolding pathway is a mixture of the previous two: strands A and G are the first to unfold, then, before the molecule completely unfolds, A refolds and F unfolds. This may happen reversibly many times in a single trajectory with consecutive folding (unfolding) of strand A and parallel unfolding (folding) of strand F. Such trajectories will be labeled \textit{mixed AG-GF} because the molecule is fluctuating between two different intermediates (\textit{AG} and \textit{GF}). These intermediates have almost the same end-to-end length, and therefore cannot be distinguished in a simple free energy landscape, as illustrated in Fig.\ \ref{fig:Elandscape}, where a single, broad minimum is observed at $L\simeq13$ nm. 

\begin{figure*}[htbp]
\centering
\subfloat[][\emph{Unfolding pathway}: AG]
{\includegraphics[angle=270 , width=0.45\textwidth]{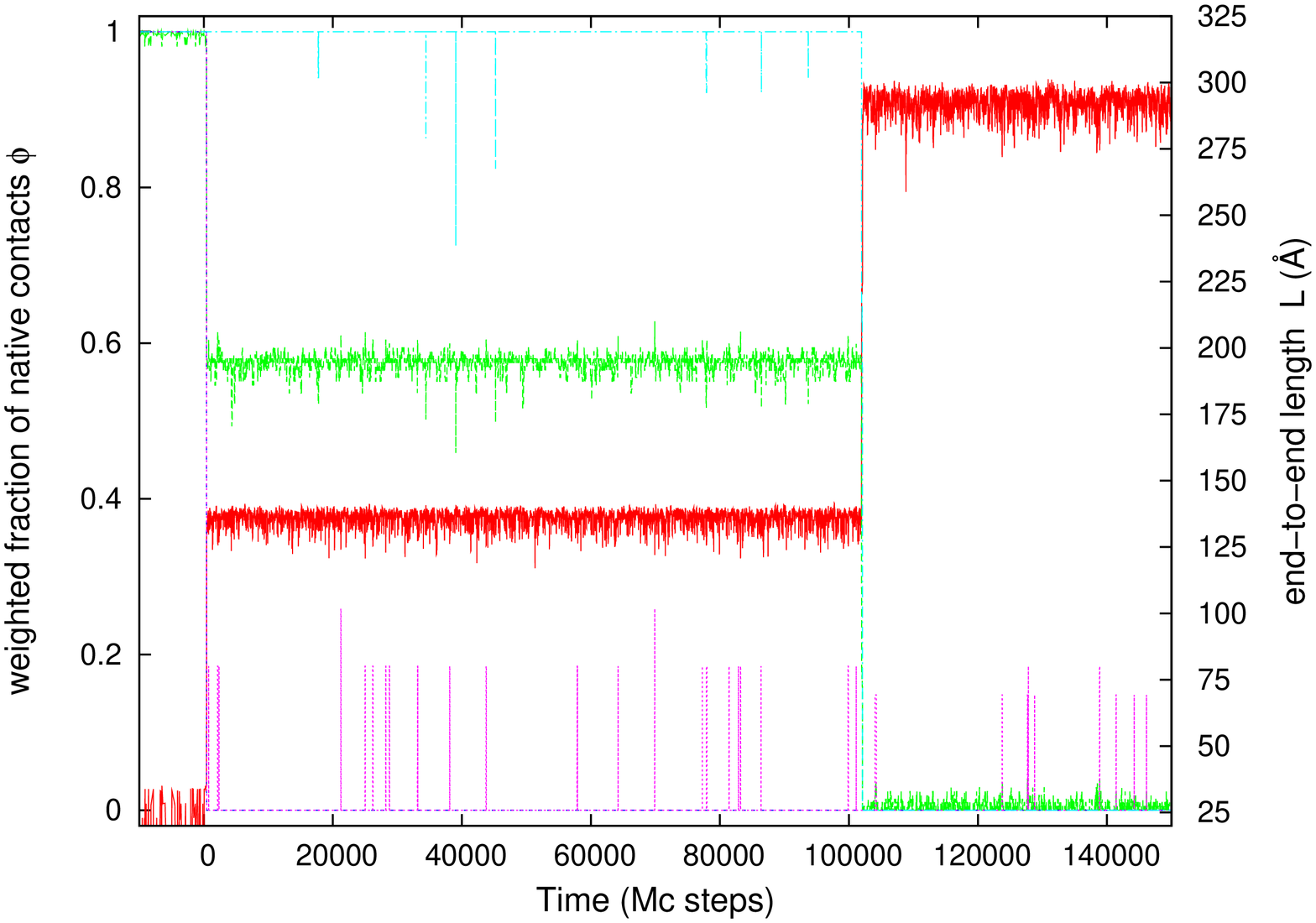}}
\subfloat[][\emph{Unfolding pathway}: GF]
{\includegraphics[angle=270 , width=0.45\textwidth]{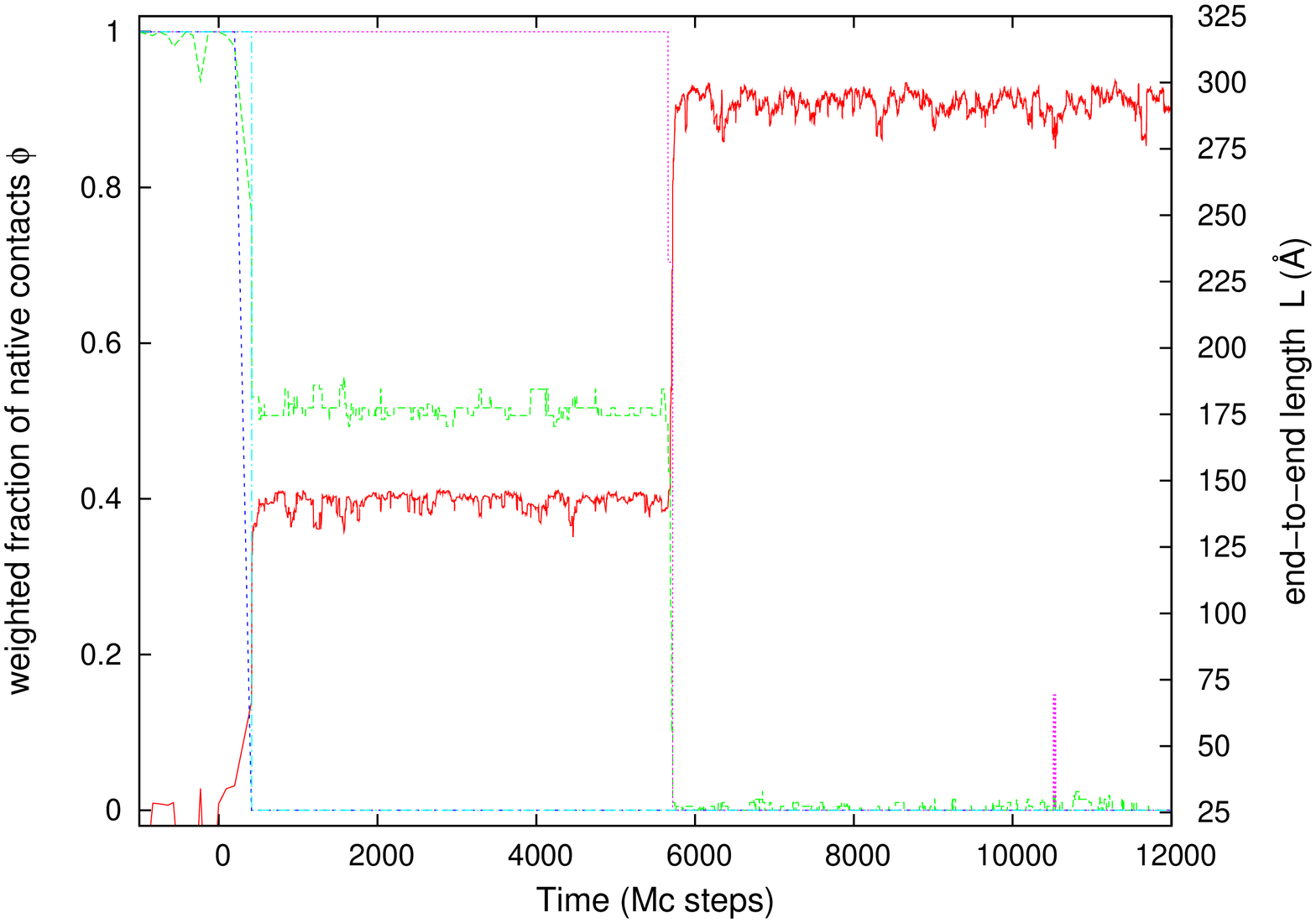}}\\
\caption{Typical MC trajectories: end-to-end length (red line) and a few order parameters as functions of time, with a $f= 65$ pN.
Green line: fraction of native contacts, whole $FnIII_{10}$. Blue line: fraction of native contacts between strands G and F.
Purple line: fraction of native contacts between strands A and B. Cyan line: fraction of native contacts between strands C and F.}
\label{fig:figureF65}
\end{figure*}

\begin{figure*}[htbp]
\centering
\includegraphics[angle=270, width=0.95\textwidth]{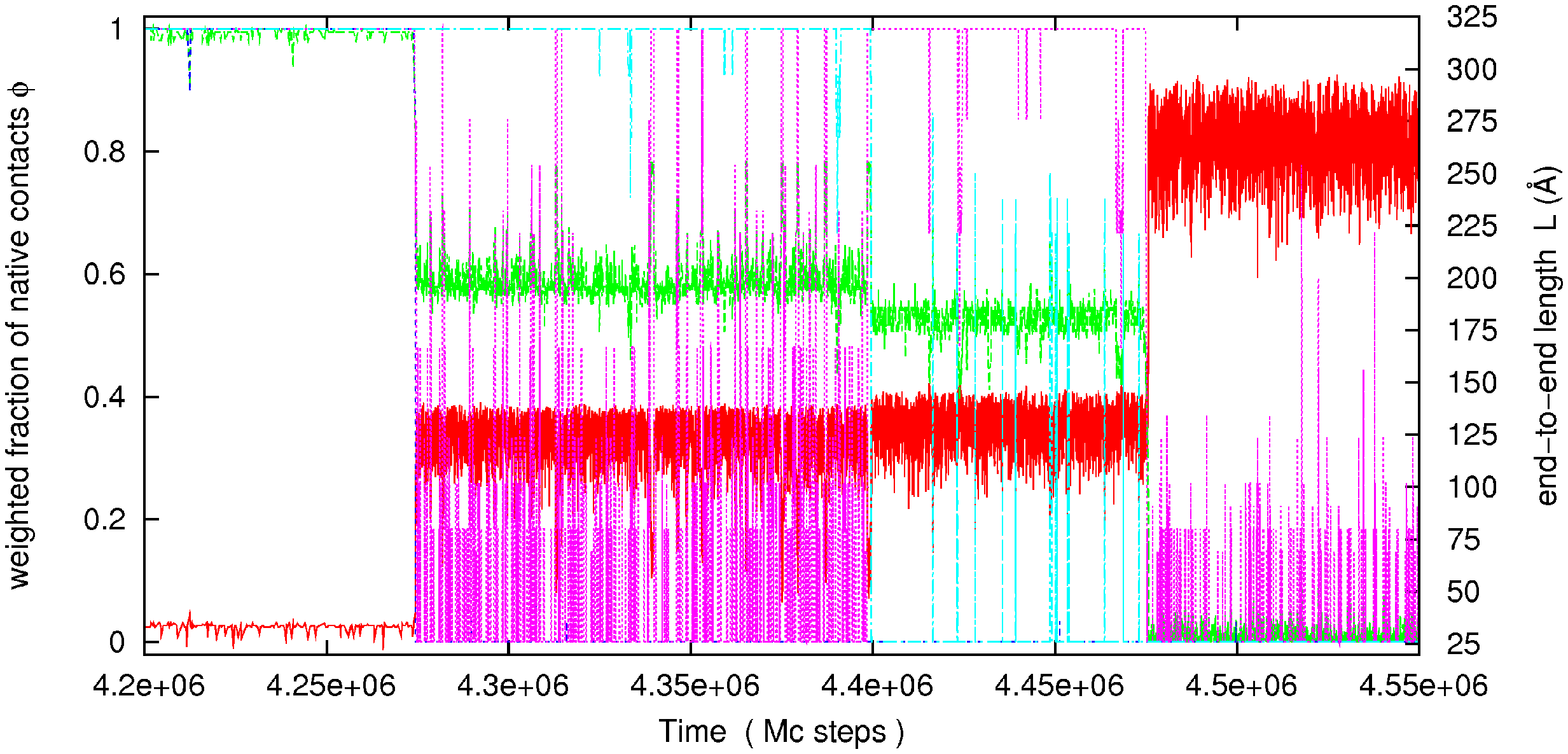}
\caption{Mixed AG-GF trajectory: MC time evolution of the end-to-end length (red line) and of some order parameters with a constant force of 28 pN. Colors as in Fig.\ \protect\ref{fig:figureF65}}
\label{fig:figureF28AGF}
\end{figure*}

Fig.\ \ref{fig:figureF65} shows two typical trajectories at $65$ pN constant force. Other typical trajectories are plotted in \cite{epaps}. During thermalization, before turning the force on at time $t=0$, the length of the polypeptide chain fluctuates around $L=0$ (Fig.\ref{fig:figureF65}a), since different orientations of the molecule are equally likely. Then, at time $t=0$, a waiting phase starts, which can be easily seen in Fig.\ref{fig:figureF65}b and Fig.\ref{fig:figureF65}d. This waiting phase corresponds to a metastable state which is characterised by an end-to-end length $\sim 3.5$ nm corresponding to the elongation in the native state.  The rise in the end-to-end length to the intermediate value is always associated to the drop in order parameters connected to two different pairs of strands, with the GF pair always involved. The order parameters which have not been plotted go to zero only when the protein reaches the fully elongated configuration (end-to-end length $\sim 29$ nm).

In Fig.\ref{fig:figureF28AGF} we report a \textit{mixed AG-GF} trajectory obtained at force $f = 28$ pN, slightly larger than the equilibrium unfolding force: after the long waiting phase there are two different intermediate states before the complete unfolding. Despite the large fluctuations in the order parameters associated to the pairs C-F and A-B, it is still possible to see their general behaviour and to recognize the first intermediate state as \textit{AG} and the second as \textit{GF}. We stress that the \textit{GF} and \textit{AG} intermediates have very similar end-to-end length and fraction of native contacts $\phi_s$ (for the whole chain), making them indistinguishable in simple, one--dimensional, free energy landscapes: indeed, they are lumped together in the broad minimum at $L\simeq13$ nm in Fig.\ \ref{fig:Elandscape}.

Both the waiting and intermediate states (as the whole unfolding process) are characterised by time lengths varying in a wide range of values for different applied forces and, because of stochasticity, for different trajectories. In Table \ref{tab:lifetime} we reported the mean life times at various constant forces. The times $\tau_{AG}$ and $\tau_{GF}$ are obtained by an average of the times occurring between the first and the second jump in the end-to-end length, which have been fixed using the respective threshold values $L=75$\AA\ and $L=225$\AA. 
These averages have been calculated only for those trajectories which exhibit the corresponding unfolding pathway, while $\tau_{AG}$ and $\tau_{GF}$ at force $f=28$ pN and $\tau_{GF}$ at force $f=40$ pN are not reported in the table because of vanishing frequencies of the corresponding trajectories, as shown in Table \ref{tab:frequencies}. 
The mean waiting phase time $\tau_{ws}$ is the average over the $100$ trajectories of the time  at which the end-to-end length  becomes longer than the threshold value $L=75$\AA. For forces $f=98$ and $122$ pN it does not make sense to define a waiting phase life time, since the protein starts to unravel as soon as the external force is applied at $t=0$. Finally, the unfolding mean time $\tau_{u}$ is the average on all the trajectories of the unfolding time, i.e. the time at which the molecule reaches the unfolding length previously defined.

\begin{figure}[htbp]
\centering
\includegraphics[angle=270, width=0.45\textwidth]{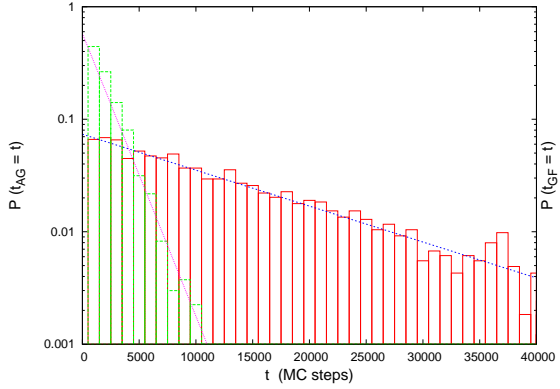}
\caption{Histograms of the intermediate life times for AG pathway (red line) and GF pathway (green line) at force $f=81$ pN. Data obtained from $3600$ different trajectories. The lines are exponential fits.}
\label{fig:istolifetimeint}
\end{figure}

\begin{figure}[htbp]
\centering
\includegraphics[angle=270, width=0.45\textwidth]{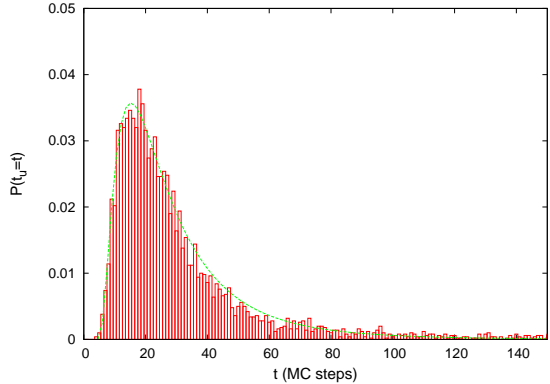}
\caption{Histograms of the unfolding times at force $f=150$ pN. Data obtained from $5000$ different trajectories and the bin size of the histogram is $1$. The fit is to a lognormal distribution.}
\label{fig:lognormalditr}
\end{figure}

The probability distributions of intermediate life times for $f=81$ pN have been plotted in Fig.\ \ref{fig:istolifetimeint}, where it can be seen that both distributions can be fitted to the negative exponential function $P(t_s) = \frac{1}{\tau_s} \exp \left\lbrace \frac{-t_s}{\tau_s} \right\rbrace$ (where $s$ is AG or GF, $t_s$ is the intermediate life time of $s$, $\tau_s = \left\langle  t_s \right\rangle $ its average), and that  \textit{AG} has a longer life than \textit{GF}. Being the unfolding time the sum of the waiting phase time and of the intermediate state time we can naively conclude that if the protein follows the GF pathway, it will reach the unfolded state earlier. For the same reason and since at $f=81$ pN the dominant contribution to the unfolding time comes from $ \tau_{GF}$ and $ \tau_{AG}$ we can argue that at this force the exponential function fits well the unfolding times distribution too \cite{JPhysCondMatter18,JCHEMPHYS127}. Furthermore, at very high forces a lognormal distribution of unfolding times has been proposed \cite{JPhysCondMatter18}. Fig.\ref{fig:lognormalditr} shows this behaviour at force $f=150$ pN and the corresponding fit to $P(t_u) = \frac{1}{\sqrt{2 \pi} \sigma \left( t_u - t_0 \right) } \exp  \left\lbrace   -\frac{\ln^2 \left( \frac{t_u - t_0}{m} \right) }{2 \sigma^2}   \right\rbrace$.

Looking at data in Table \ref{tab:lifetime} we can try to interpret the three different force ranges in Fig.\ \ref{fig:unftime_vs_force}. In the highest force range there is neither an intermediate state, nor a waiting phase and the unfolding time corresponds mainly to the MC time needed for completing the unfolding where every MC move that unravels the molecule and thus increases the length is accepted and every move that reduces the length is refused, that is, an extremely biased random walk, corresponding to the scenario proposed in \cite{AlbertoPRERC}. Lowering the force the contributions of $ \tau_{GF}$ and $ \tau_{AG}$ to the global unfolding time become important while the waiting phase, if it exists, is still quite short. Finally, in the lowest force interval, also the waiting phase gives its contribution and this matches with the larger slope of the fit line.

\begin{table}
\caption{\label{tab:lifetime} Unfolding time ($\tau_u$), waiting phase life time ($\tau_{ws}$), AG intermediate life time ($\tau_{AG}$) and GF intermediate life time ($\tau_{GF}$) at different constant forces. Values are in MC steps and are approximated averages on $100$ different trajectories at each force.}
\begin{ruledtabular}
\begin{tabular}{c|cccc}

         & $\tau_u$         & $\tau_{ws}$      & $\tau_{AG}$ & $\tau_{GF}$       \\
\hline
$28$ pN  & $2.2 \cdot 10^7$ & $8.7 \cdot 10^6$ &                  &                \\
$36$ pN  & $3.0 \cdot 10^6$ & $8.8 \cdot 10^5$ & $4.3 \cdot 10^5$ &                 \\
$40$ pN  & $8.8 \cdot 10^5$ & $2.8 \cdot 10^5$ & $6.1 \cdot 10^5$ &                  \\
$46$ pN  & $2.9 \cdot 10^5$ & $6.9 \cdot 10^4$ & $2.4 \cdot 10^5$ & $1.4 \cdot 10^4$  \\
$53$ pN  & $1.1 \cdot 10^5$ & $2.2 \cdot 10^4$ & $1.2 \cdot 10^5$ & $9.4 \cdot 10^3$   \\
$65$ pN  & $2.7 \cdot 10^4$ & $3.6 \cdot 10^3$ & $3.8 \cdot 10^4$ & $2.9 \cdot 10^3$    \\
$81$ pN  & $6.6 \cdot 10^3$ & $1.1 \cdot 10^2$ & $1.5 \cdot 10^4$ & $1.2 \cdot 10^3$     \\
$98$ pN  & $1.9 \cdot 10^3$ &                  & $4.1 \cdot 10^3$ & $7.4 \cdot 10^2$      \\
$122$ pN & $1.5 \cdot 10^2$ &                  & $5.7 \cdot 10^2$ & $1.9 \cdot 10^2$       \\
\end{tabular}
\end{ruledtabular}
\end{table}

\begin{table}
\caption{\label{tab:frequencies} Relative frequencies of unfolding pathways at constant force. $100$ trajectories for each value of the force.}
\begin{ruledtabular}
\begin{tabular}{c|cccc}

         & AG      & GF     & no intermediates & mixed AG-GF  \\
\hline
$28$ pN  & $0$     & $0$    &     $0$          & $1$       \\
$36$ pN  & $0.07$  & $0$    &     $0$          & $0.93$     \\
$40$ pN  & $0.96$  & $0$    &     $0$          & $0.04$      \\
$46$ pN  & $0.93$  & $0.07$ &     $0$          & $0$          \\
$53$ pN  & $0.67$  & $0.32$ &     $0.01$       & $0$           \\
$65$ pN  & $0.59$  & $0.31$ &     $0.1$        & $0$            \\
$81$ pN  & $0.41$  & $0.34$ &     $0.25$       & $0$             \\
$98$ pN  & $0.43$  & $0.23$ &     $0.34$       & $0$              \\
$122$ pN & $0.2$   & $0.06$ &     $0.74$       & $0$               \\
\end{tabular}
\end{ruledtabular}
\end{table}

Table \ref{tab:frequencies} shows the frequencies of various unfolding pathways. Predictably, as the force increases, the trajectories without any intermediate state become dominant and we expect them to be the only escape route at even  higher forces, as already observed in previous all--atom simulations \cite{BiophysJ96}.

%, but is at odds with simple analytical considerations of \cite{JMB323}.

%Another interesting feature to consider is the relative occurrence of \textit{AG} and \textit{GF} intermediates. The ratio of their frequencies $\nu_{AG} / \nu_{GF}$ has a minimum at force $f=81$ pN and increases with force in agreement with the idea\cite{BiophysJ96} that the unfolding pathway becomes more deterministic with increasing force. At $f=40 $ pN we didn't get any \textit{GF} trajectory.
At $f=28 $ pN, because of long life times and great fluctuations, all the trajectories are of \textit{mixed AG-GF} type. Furthermore, at such a low force,  the molecule can completely refold after it partially unravelled. This can happen many times before complete unfolding \cite{epaps}.

\subsection{Constant velocity}

We run MC simulations at six different pulling velocities ($0.03$, $0.05$, $0.1$, $0.3$, $0.5$, $1$ $\mu$m/s) with a spring constant $k = 30$ pN/nm and an initial length $\mathcal{L}_0=32$\AA. Once again, our conditions are much closer to experimental ones than most previous simulations. In constant velocity simulations, Vogel et al \cite{ProcNatlAcadSciUSA96} used $v=$50 m/s (with a spring constant of $\sim 4$ nN/nm), Klimov and Thirumalai \cite{ProcNatlAcadSciUSA97} considered $v=6$ mm/s or faster, while experimental pulling speeds \cite{JMolB319,JMolB345} were 0.4 and 0.6 $\mu$m/s (with spring constants of 45--50 pN/nm) and in vivo pulling speeds are believed to be even smaller. Only the all--atom Monte Carlo simulations by Mitternacht et al \cite{BiophysJ96} could probe constant pulling speeds in the same range as we are considering here (with a spring constant of 37 pN/nm). 

\begin{figure}
\setlength{\unitlength}{0.8cm}
\begin{center}
\begin{picture}(7,9)(0,0)
\thicklines
\put(0,0){\makebox(7,9){}}
\put(3.5,0.6){\circle{1.5}}
\put(3.5,7.8){\circle{1.5}}

\multiput(0.5,4.8)(0.3,0){4}{\line(1,0){0.2}}
\multiput(0.5,4.8)(0,0.3){4}{\line(0,1){0.2}}
\multiput(1.7,6)(-0.3,0){4}{\line(-1,0){0.2}}
\multiput(1.7,6)(0,-0.3){4}{\line(0,-1){0.2}}
\multiput(5.3,4.8)(0.3,0){4}{\line(1,0){0.2}}
\multiput(5.3,4.8)(0,0.3){4}{\line(0,1){0.2}}
\multiput(6.5,6)(-0.3,0){4}{\line(-1,0){0.2}}
\multiput(6.5,6)(0,-0.3){4}{\line(0,-1){0.2}}

\put(0.5,2.4){\line(1,0){1.2}}
\put(0.5,2.4){\line(0,1){1.2}}
\put(1.7,3.6){\line(0,-1){1.2}}
\put(1.7,3.6){\line(-1,0){1.2}}
\put(5.3,2.4){\line(1,0){1.2}}
\put(5.3,2.4){\line(0,1){1.2}}
\put(6.5,3.6){\line(0,-1){1.2}}
\put(6.5,3.6){\line(-1,0){1.2}}

\put(3.11,0.4){\begin{large} U \end{large}}
\put(3.11,7.65){\begin{large} N \end{large}}
\put(0.4,2.8){\begin{large} GF \end{large}}
\put(5.3,2.8){\begin{large} AG \end{large}}
\put(0.69,5.2){\begin{large} A \end{large}}
\put(5.52,5.2){\begin{large} G \end{large}}

\thicklines
\put(5.9,4.7){\vector(0,-1){1}}
\put(2.9,7.1){\vector(-1,-1){1}}
\put(4.1,7.1){\vector(1,-1){1}}
\put(2.0,5.4){\vector(1,0){3.1}}
\put(1.9,2.3){\vector(1,-1){1}}
\put(5.1,2.3){\vector(-1,-1){1}}
\put(2.0,3.0){\vector(1,0){3.1}}
\put(5.1,3.0){\vector(-1,0){3.1}}
\end{picture}
\end{center}
\setlength{\unitlength}{1cm}
\caption{Unfolding pathways scheme of $FnIII_{10}$ pulled at constant velocity. Intermediate states in the full square boxes have a rupture force remarkably higher than those in dashed boxes.}
\label{fig:unfpathconstvel}
\end{figure}
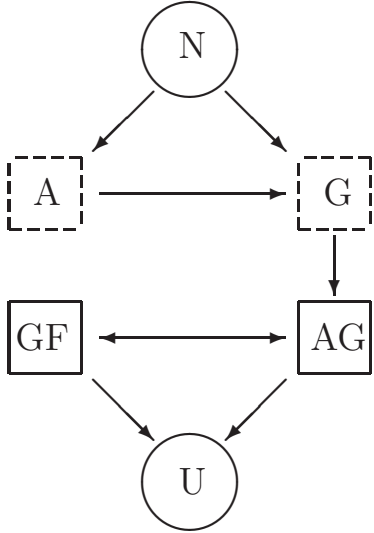

\begin{figure*}[htbp]
\centering
\subfloat[][\emph{Unfolding pathway}: G $\rightarrow$ AG]
{\includegraphics[width=0.9\textwidth]{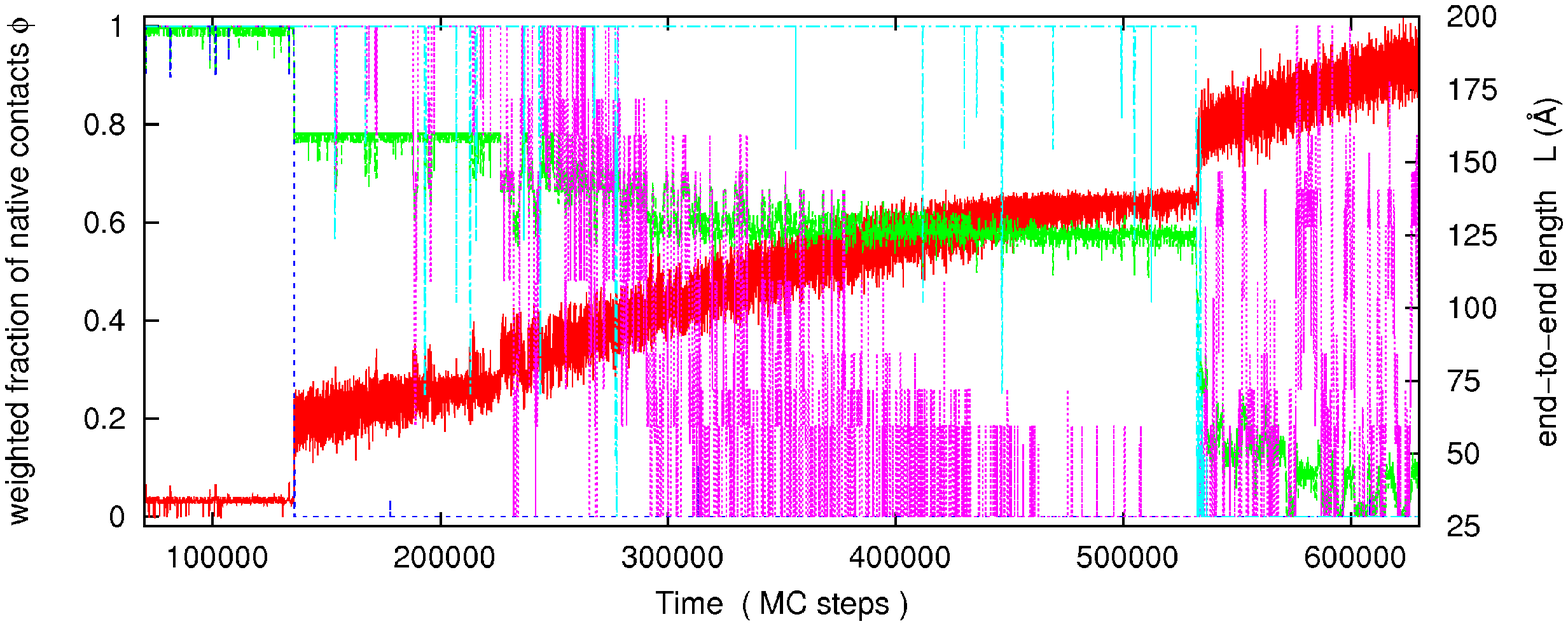}}\\
\subfloat[][\emph{Unfolding pathway}: G $\rightarrow$ mixed AG-GF (partial)]
{\includegraphics[width=0.9\textwidth]{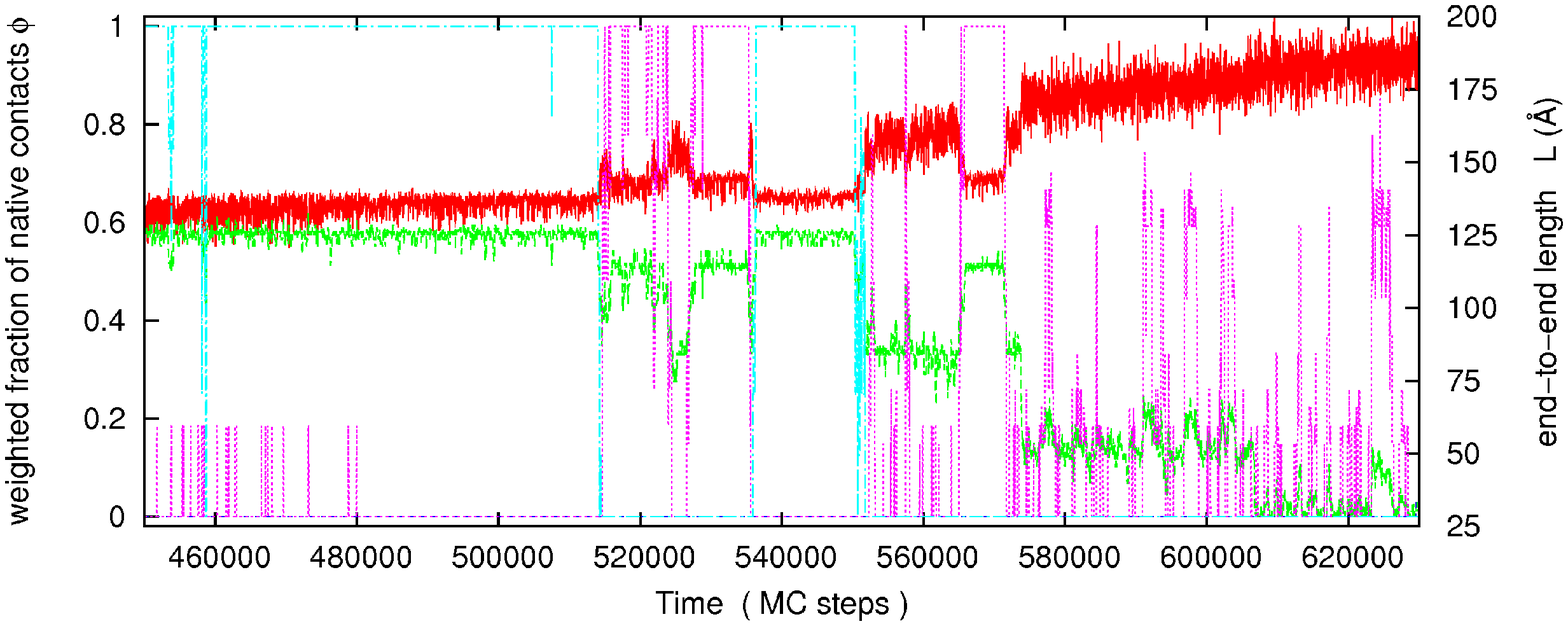}}\\
\caption{MC time evolution of the end-to-end length (red line) and of a few order parameters with a constant velocity of 1 $\mu$m/s.
Green line: weighted fraction of native contacts, whole $FnIII_{10}$. Blue line:  weighted fraction of native contacts between strands G and F.
Purple line: weighted fraction of native contacts between strands A and B. Cyan line: weighted fraction of native contacts between strands C and F.}
\label{fig:figurev25}
\end{figure*}

\begin{figure*}[htbp]
\centering
\includegraphics[width=1.\textwidth]{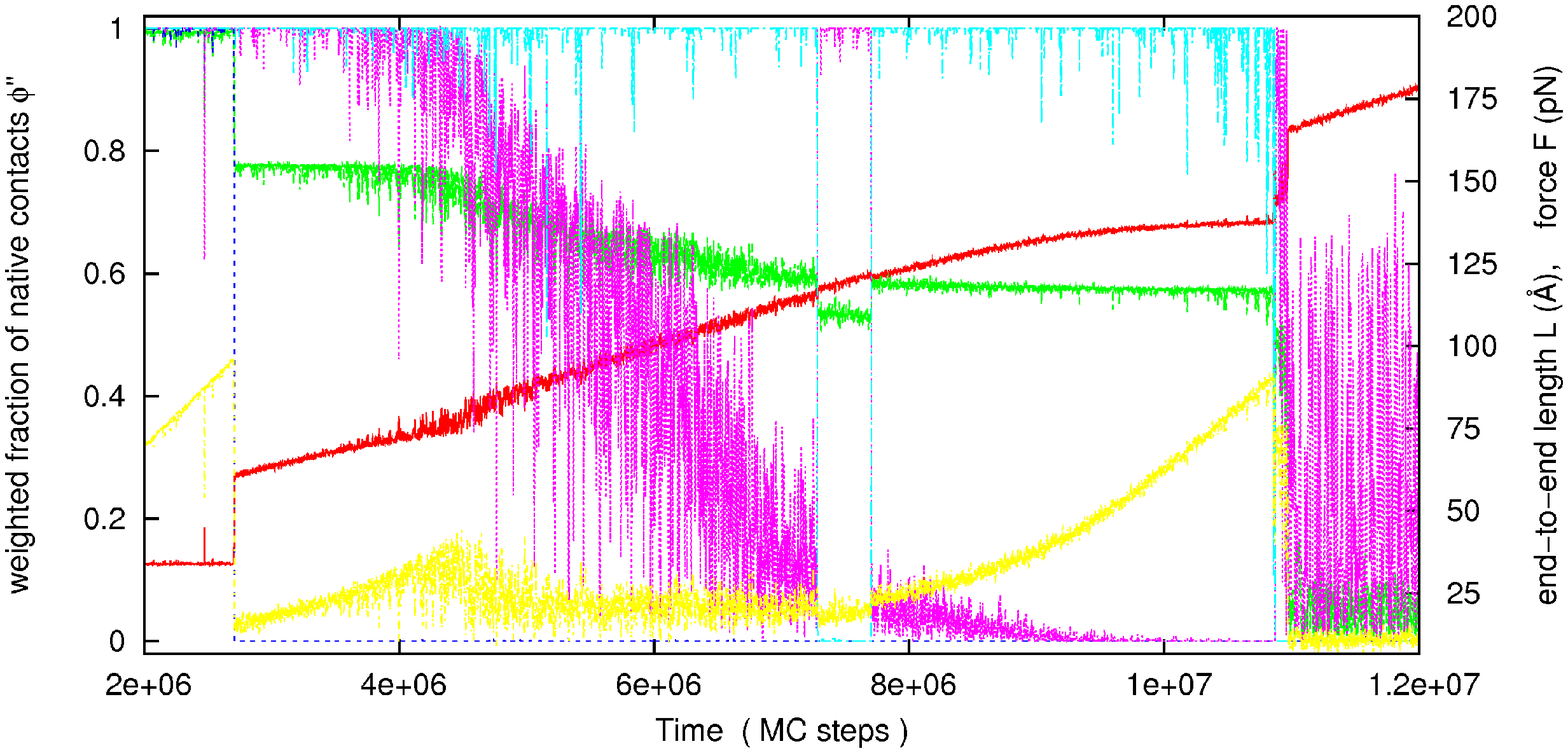}
\caption{MC time evolution of end-to-end length (red line),force (yellow line) and of some order parameters with a constant velocity of $0.03\, \mu$m/s for a mixed AG-GF trajectory.
Green line: weighted fraction of native contacts, whole $FnIII_{10}$. Blue line:  weighted fraction of native contacts between strands G and F.
Purple line: weighted fraction of native contacts between strands A and B. Cyan line: weighted fraction of native contacts between strands C and F. Bins of $2000$ MC steps have been used to reduce fluctuations in the plot.}
\label{fig:figurev125}
\end{figure*}

In Fig.\ref{fig:unfpathconstvel} we sketch the possible unfolding pathways scheme in the constant velocity case. Consistent with our constant force results and with previous simulations \cite{ProcNatlAcadSciUSA96,ProcNatlAcadSciUSA97,BiophysJ96}, at each value of $v$ most of the trajectories start with the detachment of strand G, giving rise to an intermediate corresponding to the shallow minimum around 6 nm in Fig.\ \ref{fig:Elandscape}. In few runs strand A is the first to unravel, but then it refolds, with the consequent detachment of G. 
Then the unfolding continues through a phase in which strand A is gradually unzipped and when this unzipping is completed the molecule reaches the intermediate AG (end--to--end length $\sim 13.5$ nm). Again we found a mixed AG-GF behaviour: some trajectories do not stay in the AG intermediate till the complete unfolding but they may jump from AG to GF intermediate (end--to--end length $\sim 14$ nm) and back. Table \ref{tab:freqconstvel} reports the relative frequencies of various unfolding pathways. It is worth noting that, since statistical fluctuations are greater at low pulling rates, the number of mixed AG-GF trajectories and the number of trajectories in which strand A unravels before strand G grows as pulling velocity decreases. Typical trajectories are reported in Fig.\ \ref{fig:figurev25}, in Fig.\ \ref{fig:figurev125}, and in \cite{epaps}.

The average rupture force of the native state depends on the pulling rate and is reported in Table \ref{tab:forzeconstvel}. At the pulling speed considered, the average rupture force we obtained for the native state ranges between 80 to 100 pN, which is in remarkable agreement with the AFM results. Fernandez and coworkers reported 75 pN when pulling at 0.6 $\mu$m/s \cite{JMolB319} and 100 pN at 0.4 $\mu$m/s \cite{JMolB345}. Mitternacht et al \cite{BiophysJ96} reported values from 88 pN at 0.03 $\mu$m/s to 114 pN at 0.1 $\mu$m/s. Notice that in our results the average rupture force increases with the pulling speed, as predicted by theories \cite{EvansRitchie,Evans,DudkoPNAS,AjdariBPJ2004,DudkoPRL} and verified in experiments \cite{FernandezPNAS1999}. AFM results showed a different behaviour, and this was attributed to the interactions with the other modules building up the polyprotein which is actually pulled in such AFM experiments \cite{JMolB345}. 
In the same work, the average unfolding force of the intermediate states was reported to be 50 pN. Pulling on suitable mutants, two kind of intermediates were inferred on the basis of experimental results, namely G and AB. In our model we did not observe intermediate AB, while intermediate G has an average rupture force between 40 and 50 pN. The other intermediates we observed, AG and GF, are more stable, with average unfolding forces around 70 pN.

\begin{figure}[htbp]
\centering
\includegraphics[angle=270, width=0.45\textwidth]{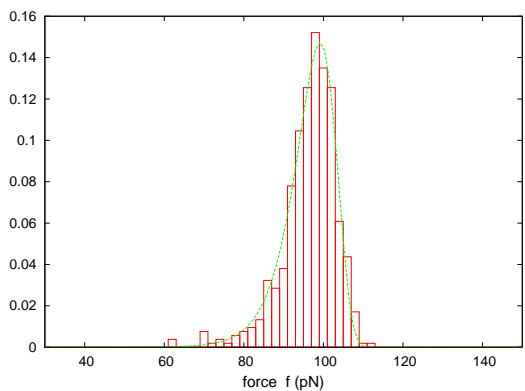}
\caption{Distributions of the rupture forces of the native state at pulling velocity $v=0.5 \, \mu$m/s. Data obtained from $500$ different trajectories; bin size of the histogram is $2$. The fit is to Eq.\ \protect{\ref{pdfEvansRitchie}}.}
\label{fig:isto_rupture_forces}
\end{figure}

The distribution of the unfolding forces is well fitted by the
theoretical result \cite{EvansRitchie} 
\begin{equation}
P(f) = \frac{1}{t_0 r}
e^{\beta f x_u} \exp \left[ -\frac{k_B T}{r x_u t_0} \left( e^{\beta f
      x_u}-1 \right) \right]. 
\label{pdfEvansRitchie}
\end{equation}
Such an equation corresponds to the rupture force probability distribution 
of a single molecular bond subject to a force that increases linearly
with a rate $r$\cite{EvansRitchie}.
In  Fig.~\ref{fig:isto_rupture_forces} we plot the unfolding force histogram at $v=0.5 \, \mu$m/s, and fit the data to eq.~(\ref{pdfEvansRitchie}), with $a=r \cdot t_0$ and $x_u$ as fitting parameters. The fit
gives $x_u = 8.0$ \AA, which is  larger than the value found for the constant force set-up, but it must be kept in mind
that the above theoretical result was derived for a force which is
linear in time with a slope $r$, while here the force is associated to the
harmonic potential which moves at constant velocity $v$.

\begin{table}
\caption{\label{tab:freqconstvel} Relative frequencies of unfolding pathways. $100$ trajectories for each value of the velocity.}
\begin{ruledtabular}
\begin{tabular}{c|cc|cc}
         & \multicolumn{2}{c|}{G}          & \multicolumn{2}{c}{A $\rightarrow$ G} \\
\hline
         & $\;$AG$\;$ & \small{mixed AG-GF}   & $\;$AG$\;$  & \small{mixed AG-GF}        \\
\hline
$1$ $\mu$m/s     & $0.82$ & $0.15$            &     $0.03$  & $0.00$             \\
$0.5$ $\mu$m/s   & $0.76$ & $0.20$            &     $0.03$  & $0.01$              \\
$0.3$ $\mu$m/s   & $0.49$ & $0.39$            &     $0.09$  & $0.03$               \\
$0.1$ $\mu$m/s   & $0.11$ & $0.85$            &     $0.01$  & $0.03$                \\
$0.05$ $\mu$m/s  & $0.08$ & $0.87$            &     $0.01$  & $0.04$                 \\
$0.03$ $\mu$m/s  & $0.07$ & $0.79$            &     $0.00$  & $0.14$                  \\
\end{tabular}
\end{ruledtabular}
\end{table}

\begin{table}
\caption{\label{tab:forzeconstvel} Average rupture forces.}
\begin{ruledtabular}
\begin{tabular}{c|c|c|c|c}
   rupture       & N $\rightarrow$ G & G $\rightarrow$ AG & AG $\rightarrow$ U & GF $\rightarrow$ U \\
\hline
$1$ $\mu$m/s     & $98.5\pm6.4$ & $40.8\pm2.6$   & $99.6\pm9.9$  & $77.3\pm7.7$ \\
$0.5$ $\mu$m/s   & $96.1\pm6.1$ & $42.2\pm2.6$   & $96.5\pm7.3$  & $77.5\pm4.0$ \\
$0.3$ $\mu$m/s   & $94.5\pm6.9$ & $43.4\pm2.4$   & $92.4\pm7.8$  & $76.5\pm5.8$ \\
$0.1$ $\mu$m/s   & $89.0\pm6.5$ & $45.4\pm1.8$   & $86.5\pm7.9$  & $69.9\pm5.8$ \\
$0.05$ $\mu$m/s  & $87.8\pm5.1$ & $46.1\pm1.6$   & $81.9\pm8.4$  & $67.6\pm5.9$ \\
$0.03$ $\mu$m/s  & $87.3\pm5.8$ & $46.9\pm1.7$   & $81.5\pm9.7$  & $66.7\pm5.4$ \\
\end{tabular}
\end{ruledtabular}
\end{table}

\section{Conclusions}
\label{Conclusions}

We have simulated constant force and constant pulling speed unfolding of $FnIII_{10}$, the tenth type III domain of fibronectin using an Ising--like model we have developed and validated in recent years, whose equilibrium thermodynamics is exactly solvable. Force and time units have been determined by comparison with existing estimates of the equilibrium unfolding force and the zero--force average unfolding time. We can probe force and speed ranges close to in vivo and experimental conditions, which was not possible in most previous simulations. 

At high enough constant force we observed two--state transitions only. At smaller forces and at all pulling speeds considered we observed several intermediates, denoted by A, G, AG and GF, based on the strands which are unfolded in each intermediate. Possible unfolding pathways are summarized in Fig.\ \ref{ClampScheme} for the constant force protocol and in Fig.\ \ref{fig:unfpathconstvel} for the constant pulling speed protocol. 

The unfolding pathways depend on the applied force or on the pulling speed, which was already observed in \cite{BiophysJ96}. Such pathways become more complex at low forces and speeds, due to the increase in fluctuations. Previous simulations and experiments showed some discrepancies in the unfolding pathways, and our work is not going to resolve such discrepancies, but some general trends are confirmed. In particular, the most frequently observed intermediate in our trajectories was AG, which was observed in all previous simulations \cite{ProcNatlAcadSciUSA96,PaciKarplus,ProcNatlAcadSciUSA97,JMB323,BiophysJ96}. In addition, constant pulling speed trajectories always visit intermediate G, which was also observed in most previous simulations \cite{ProcNatlAcadSciUSA96,ProcNatlAcadSciUSA97,JMB323,BiophysJ96} and in AFM experiments \cite{JMolB345}. On the other hand, we have never observed intermediate AB, which has been reported in many simulations \cite{PaciKarplus,JMB323,BiophysJ96} and experiments \cite{JMolB345}. We have instead observed, at low enough forces and speeds, intermediates A and GF, which were previously reported only by Gao et al \cite{JMB323} (A only) and Mitternacht et al \cite{BiophysJ96} (both A and GF). These intermediates have end--to--end lengths close to G and AG, respectively, and cannot be distinguished in the usual one--dimensional free energy landscape using the end--to--end length as a reaction coordinate. Interestingly, in our trajectories we observe fluctuations between intermediates with similar lengths, that is between A and G or between AG and GF. Fluctuations between AG and GF, in particular, are observed in most trajectories at the lowest forces and pulling speeds we have considered, and therefore one could speculate that they have some biological significance. 

>From a more quantitative point of view, given the extreme simplicity of our model, it is remarkable that many quantities we can compute agree well with the results from AFM experiments or previous simulations with similar parameters. Our estimate for the native state unfolding length is $x_u = 3.4 \pm 0.1$ \AA, to be compared with $x_u = 3.8$ \AA\ from AFM results \cite{JMolB319} and with $x_u = 4$ \AA\ from the simulations by Mitternacht et al \cite{BiophysJ96}. The average rupture force we obtained for the native state is in the range 80 to 100 pN, to be compared with results from 75 to 100 pN reported by AFM studies \cite{JMolB319,JMolB345}, and from 88 to 114 pN in the simulations by Mitternacht et al \cite{BiophysJ96}. Finally, our intermediate G has an average rupture force between 40 and 50 pN, to be compared with 50 pN found in experiments \cite{JMolB345}, though it must be mentioned that in such work the intermediate might be an average between the G and AB intermediates. 

\acknowledgments AI gratefully acknowledges financial support from
Lundbeck Fonden and from Danish Centre for Scientific Computing
(DCSC). AI is grateful to Anders Irb\"ack and to Simon Mitternacht for
helpful and stimulating discussions.

\providecommand{\noopsort}[1]{}\providecommand{\singleletter}[1]{#1}%

\newpage
\clearpage

\newcommand{\eij}{\epsilon_{ij}}
\newcommand{\dij}{\Delta_{ij}}
\renewcommand{\thefigure}{A.\arabic{figure}}

\begin{widetext}

{\Large \bf
Appendix to ``Pathways of mechanical unfolding of $FnIII_{10}$: low force intermediates''}

In this appendix we report some additional results, complementary to those reported in the main paper. 
\section{Equilibrium properties}

As mentioned in the main text, the equilibrium thermodynamics of the present model can be solved exactly. Fig.\ \ref{fig:equilibrium} shows that a sharp transition in the fraction of native bonds and in the end--to--end length of the molecule occurs upon increasing the applied external force at about 20 pN.

\begin{figure}[htbp]
\centering
\includegraphics[angle=270, width=0.45\textwidth]{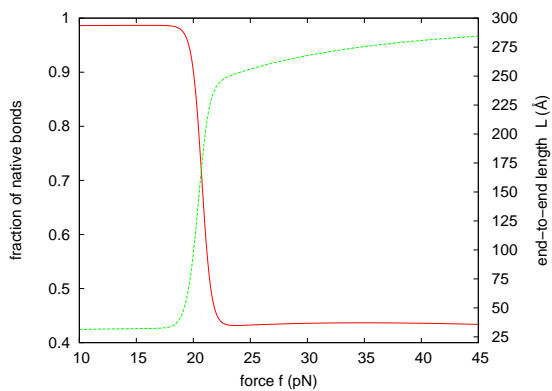}
\caption{Average fraction of native bonds (red line) and end--to--end length (green line) as a function of the external force. The temperature has been fixed to 288 K.}
\label{fig:equilibrium}
\end{figure}

\section{Unfolding pathways}

\subsection{Force Clamp}
In fig.\ \ref{fig:figureF65_app} two typical unfolding trajectories at constant force $f=65$ pN are plotted.

\begin{figure*}[htbp]
\centering
\subfloat[][\emph{Unfolding pathway}: AG]
{\includegraphics[angle=270 , width=0.45\textwidth]{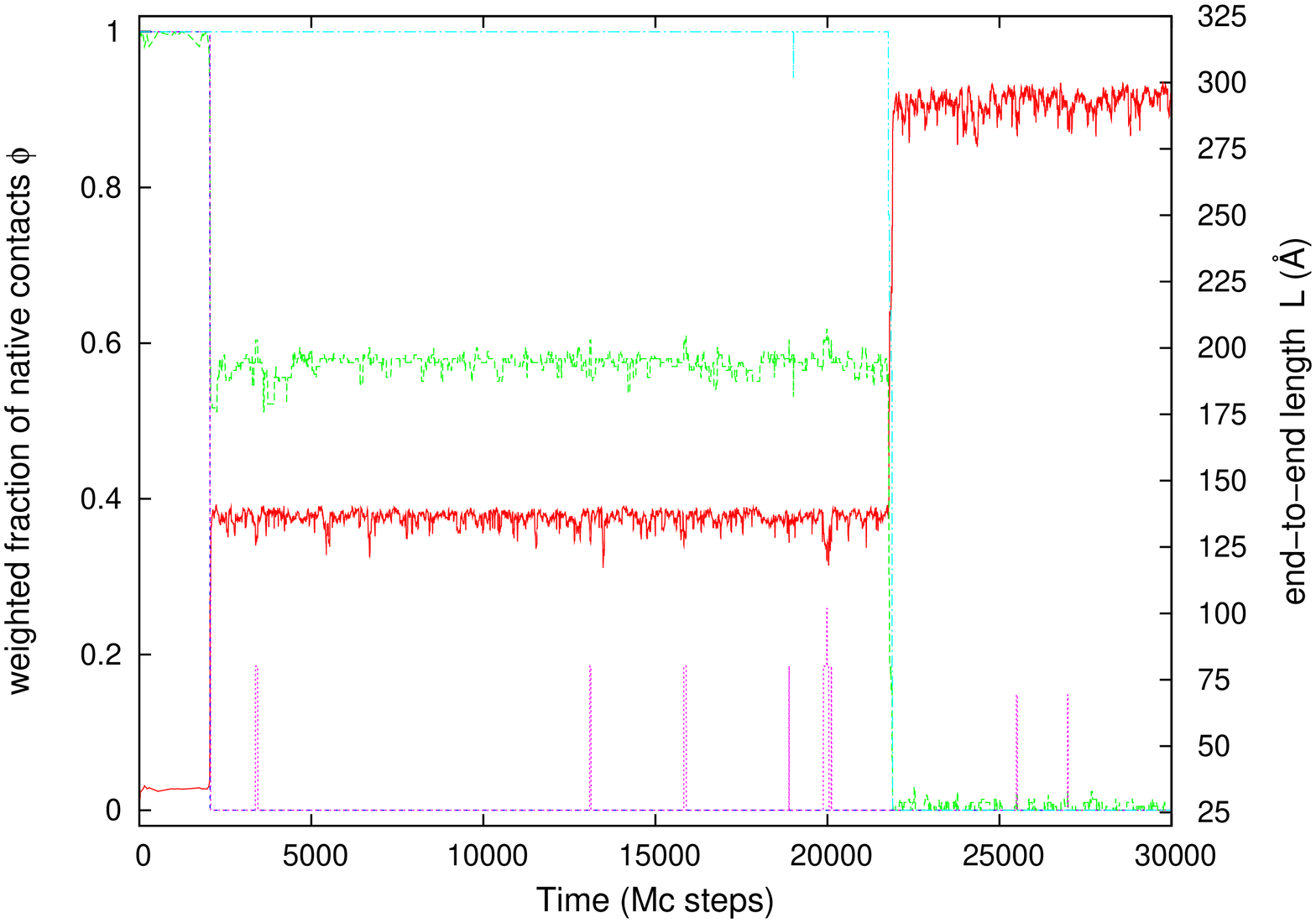}}\quad
\subfloat[][\emph{Unfolding pathway}: no intermediates]
{\includegraphics[angle=270 , width=0.45\textwidth]{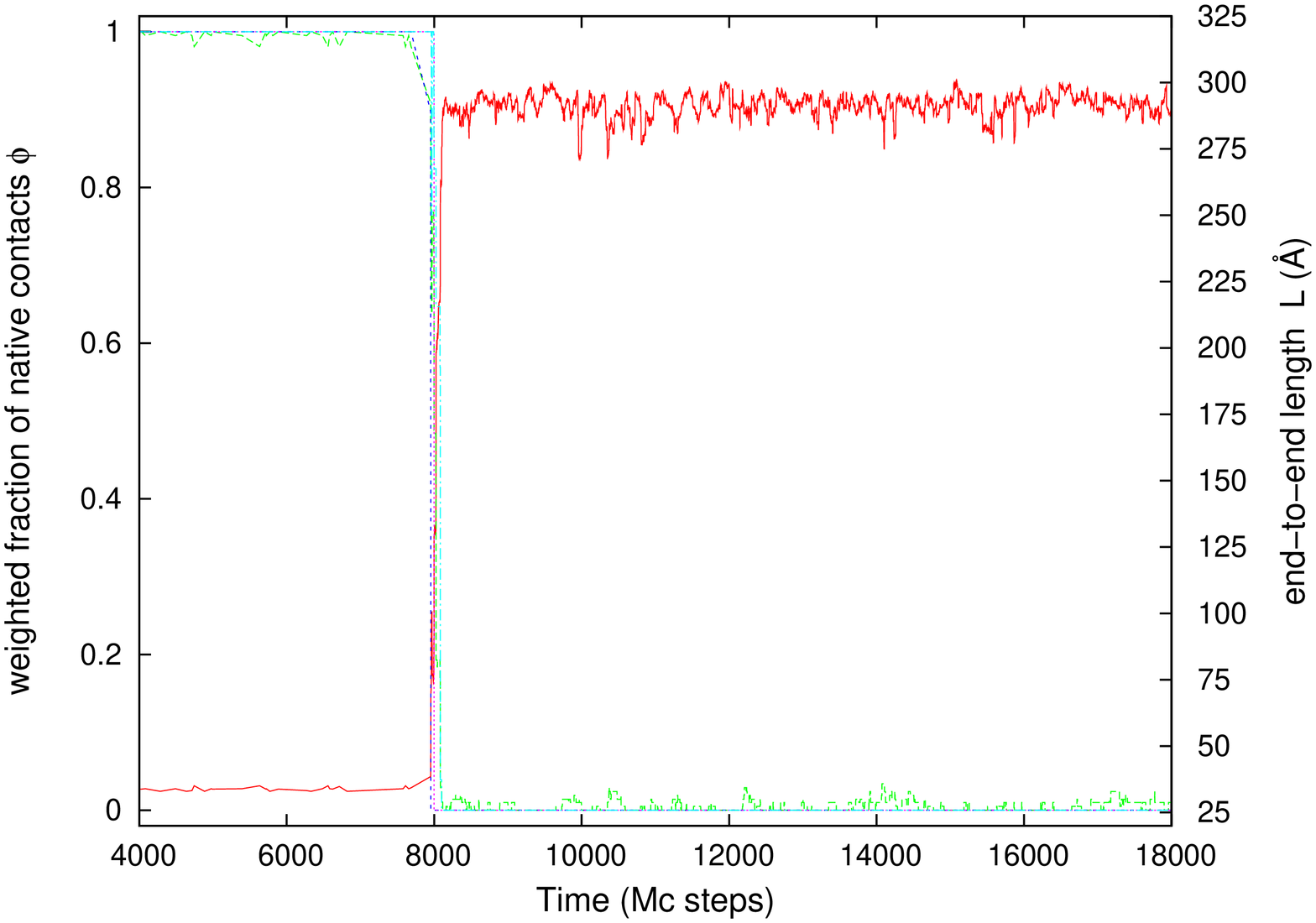}}\\
\caption{Typical MC trajectories: end-to-end length (red line) and a few order parameters as functions of time, with a $f= 65$ pN.
Green line: fraction of native contacts, whole $FnIII_{10}$. Blue line: fraction of native contacts between strands G and F.
Purple line: fraction of native contacts between strands A and B. Cyan line: fraction of native contacts between strands C and F.}
\label{fig:figureF65_app}
\end{figure*}

In fig.~\ref{fig:figureF28unf_ref} a typical unfolding trajectory at low force $f=28$ pN is plotted: the molecule hops back and forth between the folded and
the partially unfolded state, before a complete unfolding event takes place. 

\begin{figure*}[htbp]
\centering
\includegraphics[angle=270, width=0.95\textwidth]{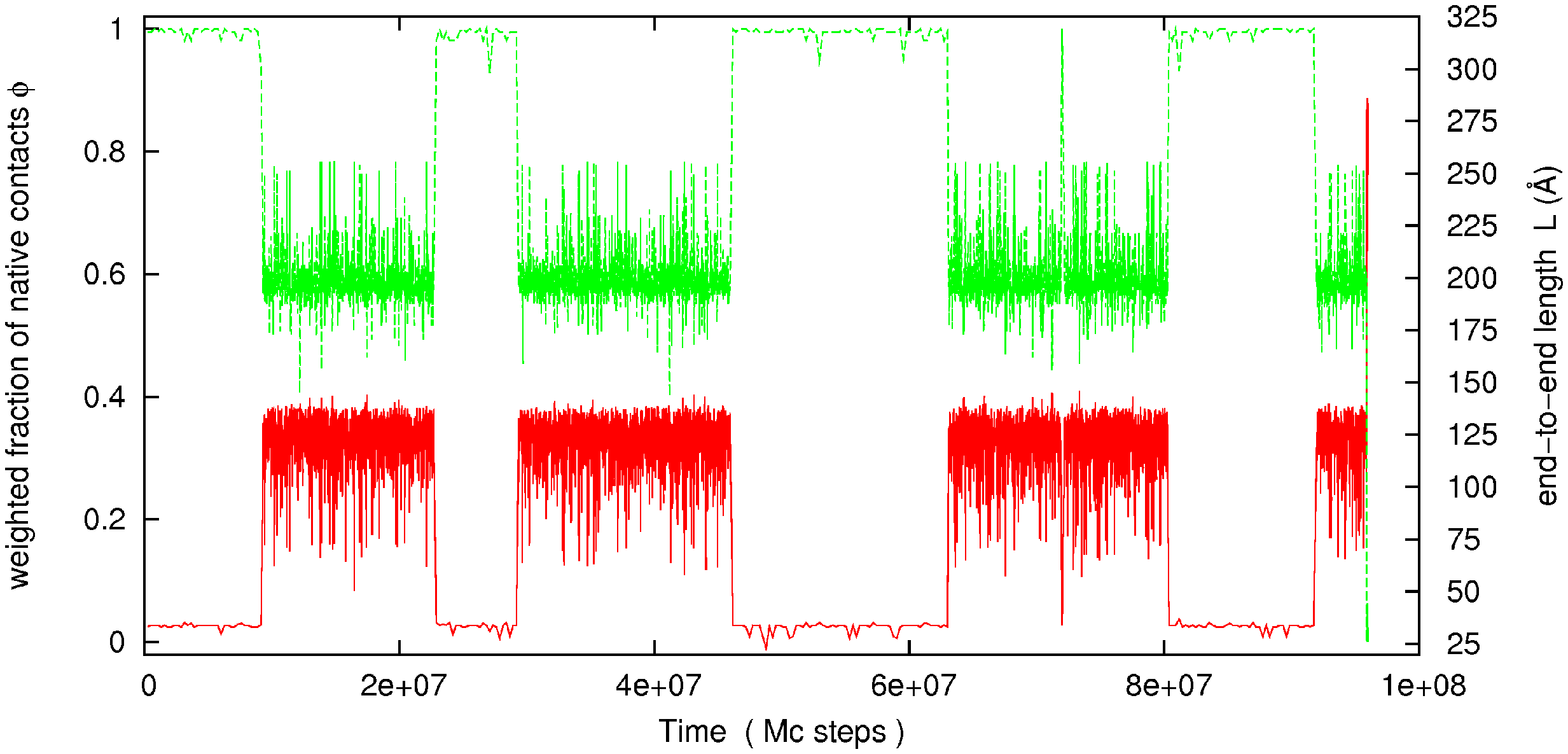}
\caption{Typical MC trajectory exhibiting sequential partial unfolding-refolding at low force $f=28$ pN:  end-to-end length (red line) and fraction of native contacts for whole $FnIII_{10}$ (green line).}
\label{fig:figureF28unf_ref}
\end{figure*}

\subsection{Constant velocity}
In fig.~(\ref{fig:figurev25_app}), we plot an
unfolding trajectory A $\rightarrow$ G $\rightarrow$ AG, for the constant velocity set-up.

\begin{figure*}[htbp]
\centering
\subfloat[][\emph{Unfolding pathway}: A $\rightarrow$ G $\rightarrow$ AG (partial)]
{\includegraphics[width=0.9\textwidth]{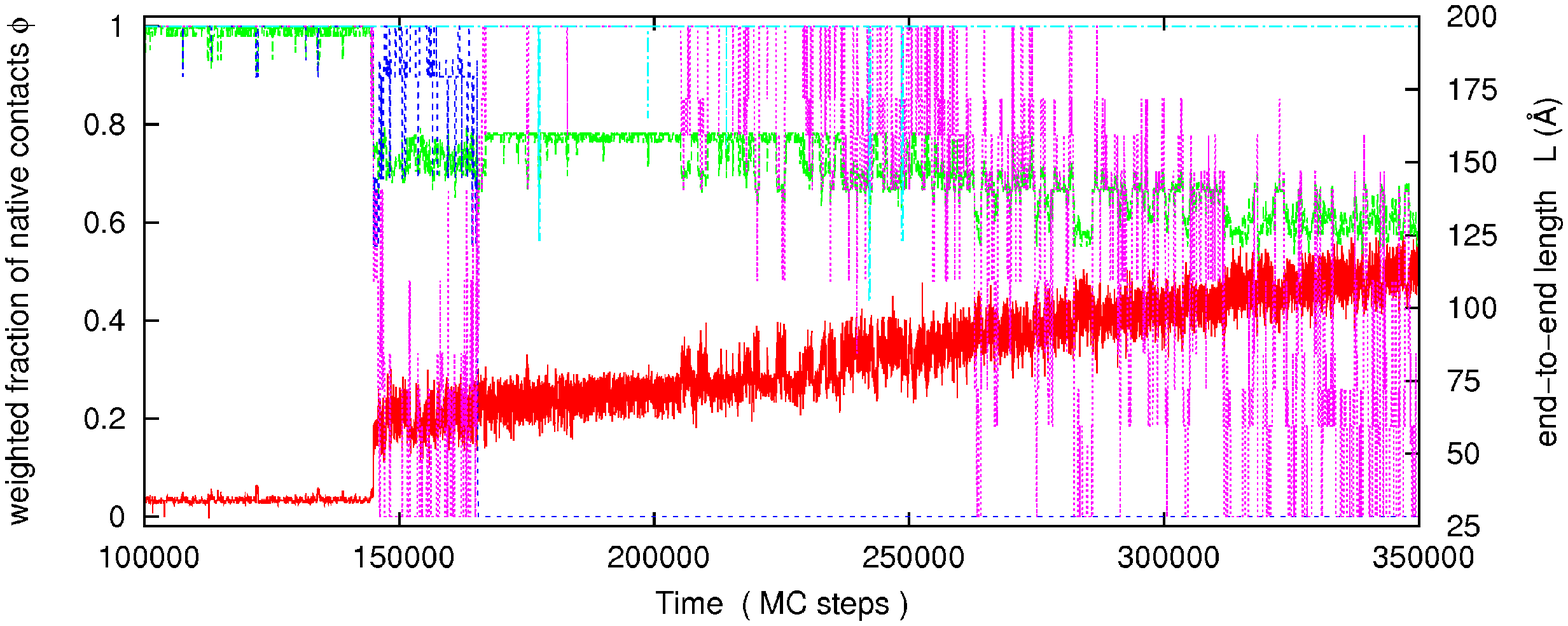}}\\
\caption{MC time evolution of the end-to-end length (red line) and of a few order parameters with a constant velocity of 1 $\mu$m/s.
Green line: weighted fraction of native contacts, whole $FnIII_{10}$. Blue line:  weighted fraction of native contacts between strands G and F.
Purple line: weighted fraction of native contacts between strands A and B. Cyan line: weighted fraction of native contacts between strands C and F.}
\label{fig:figurev25_app}
\end{figure*}

\end{widetext}


\begin{thebibliography}{39}
\bibitem[{epa()}]{epaps}
\bibinfo{note}{In the appendix at the end of this manuscript  we report additional results and figures. }
\end{thebibliography}
\end{document}